\documentstyle[prl,aps]{revtex}
\begin{document}
\widetext
 
\author{V.V. Nikolaev(a), M.E. Portnoi(a), and I. Eliashevich(b)}
\address{
(a) School of Physics, University of Exeter, Stocker Road, Exeter, EX4 4QL, UK \\
(b) GELcore LLC, 394 Elisabeth Ave., 
Somerset, NJ 08873, USA}

\title{Photon recycling white light emitting diode\\ 
based on InGaN multiple quantum well heterostructure}

\maketitle

\begin{abstract}
\noindent
A numerical method based on the transfer matrix technique is 
developed to calculate the luminescence spectra of complex layered 
structures with photon recycling. Using this method we show a 
strong dependence of the emission spectra on the optical eigenmode 
structure of the device. The enhancement of the photon recycling 
and the LED external efficiency can be achieved by placing 
the active regions inside single or coupled microcavities.
\end{abstract}

\section{Introduction} 

Recently, great interest has been shown in the use 
of light emitting diodes (LEDs) as a light source 
for illumination\cite{Nakamura}.
LEDs offer many potential advantages compared to 
conventional light sources due to their 
relatively low energy consumption, long lifetime
and high shock resistance.

Currently white-light LEDs use photo-excitation of 
phosphors to convert the blue light
from an InGaN/GaN LED into white light. 
However, phosphor has a broad emission spectrum,
thus, white LEDs based on this principle do not have 
the maximum possible luminous efficacy.

The aim of this work is to investigate the feasibility and limitations of
creating a white LED by integration within the same structure of 
several semiconductor layers emitting three basic colours.
The recent progress in the growth of InGaN-based 
double heterostructures and quantum wells
makes this alloy an ideal material for 
the LED active regions, due to the wide variety of the 
energy gaps in the InGaN system, covering 
frequencies from red to ultraviolet.  
The working regime of such a device can be achieved by electrical
pumping of active layers with the widest bandgap (blue regions),
and by making use of re-emission of the light, 
absorbed by all the active regions (so called photon recycling).  

\section{Modelling of photon recycling}

We start by writing an expression for the intensity of spontaneous emission
from a quantum well (QW) into a bulk dielectric material. 
The number of photons with energy within the interval 
[$\hbar\omega$, $~\hbar\omega \,+\, d(\hbar\omega)$], 
which are emitted from the surface $dS$ into the solid angle $d\Omega_0$ during 
the time interval $dt$ is given by  
\begin{equation}
\label{e1}
dN=W_0\,dt\,d(\hbar\omega)\,dS\,d\Omega_0=
\frac{\epsilon\omega^3 \hbar e^2p_{cv}^2g_{2D}}{2\pi^2 c^3 m_0^2}
\left[
\int_0^{+\infty}
\frac{f_hf_e\,\Gamma \, \,d\varepsilon}{(E^{2D}+\varepsilon)^2
((E^{2D}+\varepsilon-\hbar\omega)^2 + \Gamma^2)}
\right]
dt\,d(\hbar\omega)\,dS\,d\Omega_0.
\end{equation}  
Here $E^{2D}$ is the energy gap between the electron and hole 
quantized levels in the QW, $g_{2D}$ is the reduced
two-dimensional density of states, $f_e$ and $f_h$ are the electron
and hole occupation probabilities,
$\epsilon$ is the dielectric constant of the material containing the well,
$m_0$ is the free electron mass,
and $\Gamma$ accounts
for the interband relaxation and other broadening mechanisms.
The squared momentum matrix element $p_{cv}^2$ is given 
in the effective mass approximation for deep QWs
by\cite{Asada}
$$ 
p_{cv}^2=\frac{p_0^2}{2}\left(1+\gamma\right)
$$                            
for TE modes, and by
$$ 
p_{cv}^2=\frac{p_0^2}{2}\left[\left(1+\gamma\right)\cos\theta+
\left(1-\gamma\right)\sin\theta\right]
$$         
for TM modes.
Here $\theta$ is the angle between the plane-wave propagation direction 
and $Z$-axis (normal to the QW plane), and 
$\gamma=(E_{2D}-E_g)/(\hbar\omega-E_g)$,
where $E_g$ is the bandgap energy of the QW material, and $p_0$ is 
the interband momentum matrix element. 

We will use the transfer matrix technique in the plane waves basis\cite{Born} 
to analyse the optical properties of layered structures. 
The plain-wave mode interaction with the QW 
is described by a QW transfer matrix. 
If the quantum well width is much smaller than the light wavelength,
the transfer matrix has the form:
$$
\hat{M}=
\left(
\begin{array}{cc}
1+Y&Y\\
-Y&1-Y\\
\end{array}
\right).
$$
Here $Y$ is defined by the two-dimensional QW optical susceptibility 
$\chi_{2D}$: 
\begin{equation}
\label{eY}
Y=i\,\frac{2\pi k_0^2}{k_z}\chi_{2D},
\end{equation}   
where $\bf{k}$ is the wavevector of light, $k_0=\omega / c$, and 
$\chi_{2D}$ for a single-subband QW is given by\cite{Koch}            
$$
\chi_{2D}=\frac{\hbar^2{}e^2\,p_{cv}^2g_{2D}}{m_0^2}
\int_0^{+\infty}
\frac{1-f_h-f_e}
{(E^{2D}+\varepsilon)^2(E^{2D}+\varepsilon-\hbar\omega-i\Gamma)}\,
\,d\varepsilon
$$

To calculate the rate of photon extraction from a complex structure 
we consider the interference of the all possible processes, resulting 
in light emission out of the structure. These calculations require a 
knowledge of the amplitude transmission and reflection coefficients 
$t_l$ and $r_l$ for the structure part on the left of the QW, and
similar coefficients $t_r$ and $r_r$ for the structure part on the
right of the QW. We also need to know the transmission and reflection 
coefficients $t_{QW}$ and $r_{QW}$ for the quantum well itself.
Each of these coefficients can be obtained from the corresponding
transfer matrix.
  
Let us derive, for example, the power, emitted from the right side of 
the structure. 
A photon emitted inside the structure to the right can be transmitted
directly to the outside medium,
or it can be consecutively reflected from the right and left sides
of the structure and finally will be transmitted outside and so on. 
The outgoing electric field, resulting from all these processes 
is given by the sum:
$$
E_{r\to r}=t_r+r_rr_l^*t_r+r_rr_l^*r_rr_l^*t_r+..=\frac{t_r}{1-r_l^*r_r},
$$ 
where the star in $r_l^*$ indicates, that this coefficient includes 
the reflection from the emitting QW. 

Similarly, photons emitted inside the structure to the left 
can undergo multiple reflections and eventually escape from 
the structure to the right.
These processes give a second part of the external field: 
$$
E_{l\to r}=\frac{r_lt_{QW}t_r}{(1-r_l^*r_r)(1-r_{QW}r_l)}
$$
Thus, we obtain the expression for the emission from
the right side of the structure:
\begin{equation}  
\label{e2}
dN_{Er}=W_0*\left| \frac{t_r}{1-r_l^*r_r}
\left[1+\frac{r_lt_{QW}}{1-r_{QW}r_l}\right]
\right|^2
\frac{\sqrt{\epsilon_r}k_{zr}}
{\sqrt{\epsilon_0}k_{z0}}\,
dt\,d(\hbar\omega)\,dS\,d\Omega_e,  
\end{equation}
where the ratio $\sqrt{\epsilon_r}k_{zr}/\sqrt{\epsilon_0}k_{z0}$ 
accounts for
the change in solid angle that is due to refraction for plane 
waves. The indices $r$ and $0$ are related to the outside 
medium and the layer containing the QW, respectively.
To obtain the total density of the external light intensity we have 
to sum over all QWs and integrate Eq.(\ref{e2})
over the external solid angle $\Omega_e$.

If we neglect the reflection from the QW by  
substituting $t_{QW}=1$ and $r_{QW}=0$ into Eq.(\ref{e2}),
the formula for extraction becomes analogous to one 
obtained using the source-term method\cite{Benisty}. 
 
For the quantitative description of the recycling process, 
we have to calculate the rate of absorption by the quantum well 
${\rm QW_a}$ of the spontaneous emission from the other well,  
${\rm QW_e}$. 
Thus, we need to know the induced electric field at the position 
of ${\rm QW_a}$.  
The power flux balance shows, that the rate of absorption 
of emitted photons
by the unit area of surface of ${\rm QW_a}$ is given by:
\begin{equation}
\label{eNEW}
W_{a}=-\frac{1}{2}\sqrt{\epsilon}k_z|E|^2\,{\rm Re}(Y_{QW_a})\,W_{0QW_e},
\end{equation}
where $E$ is the complex amplitude of the field at the ${\rm QW_a}$ 
location, $Y_{QW_a}$ is defined by Eq.(\ref{eY}), and $W_{0QW_e}$
is defined by Eq.(\ref{e1}). 
The induced field can be calculated in a similar 
fashion as it was done for extraction:  
$$
E=\frac{t(1+r_{al}^*)}
{(1-r^*r_{er}^*)(1-r_{al}^*r_{ar})}
\left( 1+
\frac{t_{QW_e}r_{er}}{1-r_{er}r_{QW_e}}\right)
,
$$  
where $r_{al}$ ($r_{ar}$) 
is a reflection coefficient for the part of structure to the left 
(right) of the well ${\rm QW_a}$, $r_{er}$ is a reflection coefficient 
for all the layers to the right of the well ${\rm QW_e}$, and 
the coefficients $r$ and $t$ 
correspond to the part of the structure between ${\rm QW_e}$ and 
${\rm QW_a}$. Here we assume that ${\rm QW_a}$ is on the left of ${\rm QW_e}$.
The formula for the opposite case is similar.  

One of the channels for the photon to escape from the recycling process is 
to be absorbed in a metallic mirror. Placing such a mirror onto the left 
side of the structure, and denoting the reflection coefficient of the left 
part of the structure, excluding the mirror, as $r_l$,
the reflection coefficient from the mirror as $r_m$, 
the reflection coefficient for the wave incident
from the mirror as $r_s$, and the transmission coefficient from the QW 
to the mirror as $t$, we obtain the following expression for the number 
of absorbed photons:
\begin{equation}
\label{eM}
dN_m=W_0
\left(1-|r_m|^2\right)
\left|
\frac{t}{(1-r_lr_r^*)(1-r_mr_s)}
\left(1+\frac{r_rt_{QW}}{1-r_rr_{QW}}\right)
\right|^2
dt\,d(\hbar\omega)\,dS\,d\Omega_0.
\end{equation}

We restrict our consideration by relatively low pumping levels, 
which are typical for the diode operation regime. 
Thus, the expression for the electron density\cite{Koch} 
in the one-subband 
QW can be simplified (temperature is measured in energy units):
$$ 
n=\frac{m_c}{\hbar^2\pi}T\,\ln[1+\exp({\mu_e}/T)]\approx
\frac{m_c}{\hbar^2\pi}T\exp({\mu_e}/T),
$$ 
and the occupation probabilities can be expressed as:  
$$
f_e=\frac{1}{1+\exp(\frac{\varepsilon_e-\mu_e}{T})}
\approx\exp\left(\frac{\mu_e-\varepsilon_e}{T}\right)\approx
n\,\frac{\hbar^2 \pi }{m_cT}\,\exp(-{\varepsilon_e}/T),
$$
$$
f_h=\frac{1}{1+\exp(\frac{\varepsilon_h-\mu_h}{T})}\approx
\exp\left(\frac{\mu_e-\varepsilon_e}{T}\right)\approx
p\frac{\hbar^2\pi}{m_{hh}T}\exp(-{\varepsilon_h}/T),
$$
where $m_e$ ($m_{hh}$) and $\mu_e$ ($\mu_h$) are the electron 
(heavy hole) effective mass and quasi-Fermi-level, respectively. 
Under the low-pumping assumption we can rewrite $W_0$ in the form:
\begin{equation}
\label{e3}
W_0=
\frac{\epsilon\omega^3 \hbar e^2p_{cv}^2g_{2D}}{2\pi^2 c^3 m_0^2}
\left[
\int\limits_0^{+\infty}
\frac{\Gamma \, \exp(-\varepsilon / T)\, 
\,d\varepsilon}
{(E^{2D}+\varepsilon)^2((E^{2D}+\varepsilon-\hbar\omega)^2 +\Gamma^2)}
\right]\,np.
\end{equation}  

The total rate of light emission into the external medium can be obtained by 
substituting Eq.(\ref{e3}) in Eq.(\ref{e1}) and integrating Eq.(\ref{e1})
over the external solid angle and photon energies. 
As a result the rate of radiative recombination depends on the carrier 
densities as $R_{ext}=E\,np$.

Due to the large photon energy separation between different colours, 
the absorption rate for the short-wavelength light is independent of 
the carrier density in the narrow-gap QW.  
For the light emitted from ${\rm QW_e}$ the rate of 
absorption in ${\rm QW_a}$  is given by the expression 
$R_{abs}=A\,n_{e}p_{e}$, where $A$ is calculated by integrating  
Eq.(\ref{eNEW}) over the total solid angle of emission and over the
photon energies.
Here we assume that all QWs are embedded in layers of dielectric
material with the refractive index equal to the highest
one in the real structure.
This trick allows as to handle the interaction of active layer with 
the evanescent wave. However, introducing several sufficiently 
thin layers with high dielectric constant does not alter the optical 
properties of the structure.    
 
A similar relation holds for the rate of absorption in
a metallic mirror: $R_m=M\,np$. Coefficient $M$ is obtained by substituting
Eq.(\ref{e3}) in Eq.(\ref{eM}) and integrating the result over the 
solid angle and photon energies.
Note, that due to charge-neutrality the electron and hole densities 
in each QW are equal to each other, $n=p$.

The steady-state carrier densities are given by the balance 
between generation of electron-hole pairs in the QWs 
and their recombination, both radiative and non-radiative.
The generation processes include electric current pumping of the 
blue QWs and the re-absorption of emitted light throughout the 
structure. The recombination output goes to external emission
and internal losses, which we treat as absorption in the other QWs 
and non-radiative recombination in the given QW.    
In our approach this leads to the following equation for each QW:
$$
R_i=E_in_i^2+A_in_i^2+M_in_i^2+{n_i}/{\tau_i}, 
$$ 
where $R_i$ is the pumping rate and $\tau_i$ is the non-radiative 
recombination time.  
Assuming that $\tau_i$ does not depend on the carrier density, 
we get:
\begin{equation}
\label{e4}
n_i=\frac{\sqrt{(1/\tau_i)^2+4(A_i+E_i+M_i)R_i}-1/\tau_i}{2\,(A_i+E_i+M_i)}.
\end{equation}
Then the external efficiency for each QW can be obtained as
$$
\eta_i=\frac{E_in_i^2}{(A_i+E_i+M_i)n_i^2+n_i/\tau_i}.
$$
  
\section{Results and discussion} 
We used the numerical method described above to investigate 
a number of different types of structures.
It was revealed, that the emission spectra and external efficiencies
depend drastically on the active layer positions and on the mode 
structure of the device.
Figure 1 shows the calculated spontaneous emission for a structure, 
which represents GaN microcavity containing three QWs 
and covered on one side by a metallic mirror. 
The width of the GaN layer and the energies
of interband transitions in red, green and blue QWs are
chosen in such a way that the structure 
can be regarded as the $5\lambda/2$ resonator
for the normally propagating red-light waves, $6\lambda/2$ resonator
for the green-light waves and $7\lambda/2$ resonator for the 
blue-light waves.
We placed each QW in an antinode of a resonant mode, associated with
the QW colour.
Our calculations were performed for different times of non-radiative       
recombination ranging from 1 ns to 10 ns, and we assumed the carrier 
density in blue QW to remain constant. 
One can see, that if the internal losses are high, the recycling 
efficiency is low and a blue light only is emitted. 
When the internal efficiency increases, the emission from the optically 
pumped QWs becomes comparable with the blue-light intensity. 
The green-light intensity usually remains smaller than both blue 
and red because of the strong re-absorption in the red active region.
However, this is beneficial for the white-light generation, because 
of the high sensitivity of the human eye in the green region of 
the spectrum.
If high non-radiative loses are present, the recycling process 
can be enhanced by introducing more red and green QWs in the 
structure and by building a Bragg reflector for the blue wavelength.
A possible way to enhance the external efficiencies of all three 
colours is to place the active regions into coupled microcavities.

\newpage

\begin{figure}
\begin{center}
\includegraphics{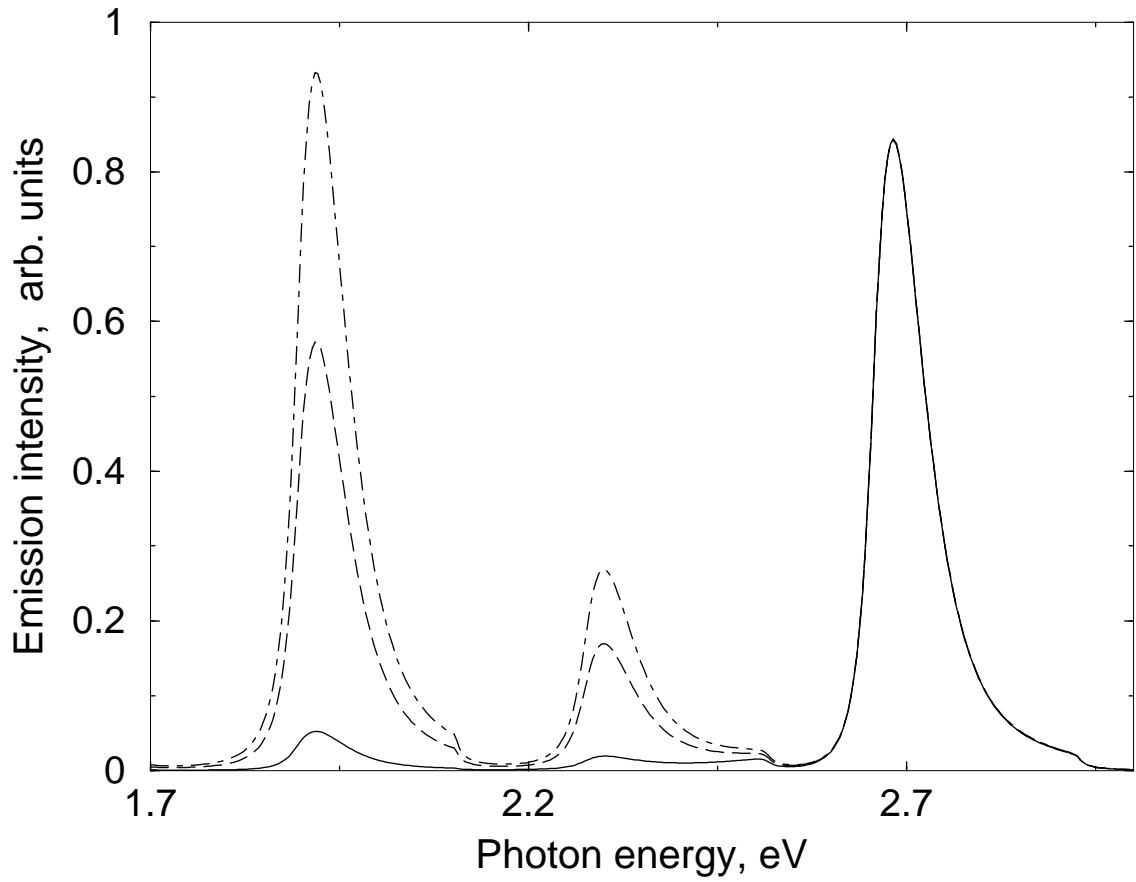}
\end{center}
\vskip 16truecm
\caption{
Spontaneous emission spectra from a microcavity, containing three quantum 
wells.
The non-radiative recombination times are 
1 ns (solid line), 5 ns (dashed line) and 10 ns (dot-dashed line).
The carrier density in the blue QW is the same for all three curves.
}
\end{figure}

\end{document}